\documentclass[prb,showpacs]{revtex4}

\usepackage{graphicx}

\begin{document}

\title{Proof of the Atmospheric Greenhouse Effect}

\author{Arthur P. Smith}
\email[E-mail:]{apsmith@aps.org}
\affiliation{American Physical Society, 1 Research Road, Ridge NY, 11961}

\begin{abstract}
A recently advanced argument against the atmospheric greenhouse
effect is refuted. A planet without an infrared absorbing atmosphere
is mathematically constrained to have an average temperature less than
or equal to the effective radiating temperature. Observed parameters
for Earth prove that without infrared absorption by the atmosphere, 
the average temperature of Earth's surface would be at least 33 K
lower than what is observed.
\end{abstract}

 \maketitle

\section{Introduction}
The results presented here are not new. However the form of presentation
is designed to clearly and accurately respond to recent
claims\cite{gerlich} that a physics-based analysis can
``falsify'' the atmospheric greenhouse effect. In fact, the standard
presentation in climatology textbooks\cite{liou} is accurate in all
material respects. The following explores in more detail certain points
that seem to have been cause for confusion.

First presented are the definitions of basic terms and the relevant
equations for the flow of energy. The situation for a planet with
no infrared-absorbing atmosphere is then examined, and a constraint
on average temperature is proved.

Several specific models of planets with no infrared-absorbing atmospehere
are then solved, including one presented by Gerlich and
Tscheuschner\cite{gerlich}, and it is verified that all satisfy this
constraint.

A simple infrared-absorbing atmospheric layer is added to these models,
and it is proved that the temperature constraint is easily violated,
as is shown by the observational data for Earth.

\section{Definitions and Basic Equations}
Define the incoming irradiance $S$ as the energy per unit area and per
unit time arriving at a planet from a stellar source. The actual
radiation field is characterized by a spectrum of wavelengths and
(depending on the size of the star(s) and distance to the planet)
a small spread in directions. $S$ is an integral over all wavelengths
and propagation directions of the radiant specific energy at the
distance of the planet from the star. As the planet moves through
its yearly orbit, the value of $S$ will vary, so it is strictly a
function $S(t)$ of time. 

The energy per unit time arriving at the planet is the product of $S(t)$
with the area the planet subtends in the plane perpendicular
to the radiant propagation direction. For a spherical planet of
radius $r$, this area is simply $\pi r^2$. So total arriving
energy (energy per unit time, or power) from space is

\begin{equation}
 E_{in}(t) = \pi r^2 S(t)
\end{equation}

For a given location ${\bf x}$ on the planetary surface, the normal to
the plane of the local surface makes an angle $\theta({\bf x}, t)$ with
the radiation propagation direction of a given stellar source. Only one
side of the planet will be lit at any given time from that source; this
can be generally indicated by those angles $\theta$ from $0$ to $\pi/2$
radians. Angles from $\pi/2$ to $\pi$ would be unlit. So local incoming
irradiance would be:

\begin{equation}
s({\bf x}, t) = \cos(\theta({\bf x}, t)) S(t) \label{incomingLocalIrradiance}
\end{equation}

Integrating this on a spherical planet (including only the lit side)
gives the $\pi r^2$ factor in equation 1.
  
Define the albedo $a$ of the planet as the fraction of incoming
irradiance that is reflected. $a$ is also a local property
(for Earth much is reflected by clouds and ice), so the locally reflected
energy is $a({\bf x}, t) s({\bf x}, t)$. Integration across the lit
side of the planet gives a well defined reflected energy:

\begin{equation}
E_{reflected}(t) = S(t) \int a({\bf x}, t) \cos(\theta({\bf x}, t)) d{\bf x}
\end{equation}

An effective albedo $a_{eff}$ can then be defined by the ratio of reflected
to incoming energy across the planet as a whole:

\begin{equation}
E_{reflected}(t) = a_{eff}(t) E_{in}(t) = \pi r^2 a_{eff} S(t)
\end{equation}

The difference between incoming and reflected energy is what the planet
absorbs (again, per unit time):

\begin{equation}
E_{absorbed}(t) = \pi r^2 (1 - a_{eff}(t)) S(t)
\end{equation}

The fundamental characteristic of a planet in space is that of no
material interchanges with its surroundings. The only substantive
way energy can come in is through electromagnetic radiation,
and the only way energy can leave is similarly through the planet's
own electromagnetic emissions. There is a very small correction
from gravitational tidal forces, and a planet also receives
a small net energy input from internal radioactive decay, but for
planets like the Earth these are thousands of times smaller than
the stellar input.

A planet with no incoming absorbed energy would reach thermodynamic
equilibrium with the cosmic microwave background, with a uniform
temperature of about 2 K. Absorption of incoming stellar irradiance
results in heating until a steady state with equal incoming and outgoing
energy (measured outside the atmosphere, and averaged over one planetary
revolution, or whatever the most important variation in time) is reached.
Define $T({\bf x}, t)$ as the local surface temperature of the planet,
and $\epsilon({\bf x}, t)$ as the local emissivity. Thermal radiation
from the surface is then given by the Stefan-Boltzmann law\cite{riedi}:

\begin{equation}
E_{emitted}(t) = \sigma \int \epsilon({\bf x}, t) T({\bf x}, t)^4 d{\bf x}
\end{equation}

Similar to the effective albedo, an effective emissivity and effective
radiative temperature can be defined as averages over the planetary surface:

\begin{equation}
T_{eff}(t)^4 = {1 \over 4 \pi r^2} \int T({\bf x}, t)^4 d{\bf x}
\end{equation}

and

\begin{equation}
\epsilon_{eff}(t) = {1 \over 4 \pi r^2 T_{eff}(t)^4} \int \epsilon({\bf x}, t) T({\bf x}, t)^4 d{\bf x}
\end{equation}

Total radiated thermal energy from the surface can then be written in
terms of the effective temperature and albedo:

\begin{equation}
E_{emitted}(t) = 4 \pi r^2 \sigma \epsilon_{eff}(t) T_{eff}(t)^4
\end{equation}

For a planet with no atmosphere, or with an atmosphere that doesn't absorb
electromagnetic radiation to any significant degree, all this surface-emitted
thermal radiation escapes directly into space, just as all the absorbed
stellar radiation reaches the ground. So the net rate of change in energy
of the planet at time t is:

\begin{equation}
\dot{E}_{planet}(t) = E_{absorbed}(t) - E_{emitted}(t) =
 \pi r^2 (1 - a_{eff}(t)) S(t) - 4 \pi r^2 \sigma \epsilon_{eff}(t) T_{eff}(t)^4
\label{planetNetEnergy}
\end{equation}

We will look at relevant constraints associated with an absorbing
atmosphere via a simple model later in the discussion.

The orbital processes for a planet (and any internal variability in
the star) determine the variation in $S(t)$;
that combined with rotation and internal dynamics gives variability
in $a_{eff}(t)$, $\epsilon_{eff}(t)$, and $T_{eff}(t)$. As a result the
planet may experience natural periods of warming or cooling
as $\dot{E}_{planet}(t)$ goes positive or negative, respectively. On
average, however, over time, this rate of energy change should
come very close to zero as long as all the input parameters are
reasonable stable over the long term. If it didn't average to zero for
a long period of time, the energy of the planet would cumulatively build
or decline.

In addition to the effective temperature obtained from averaging
temperature to the fourth power, relevant for the thermal radiation
problem, we should also look at the more natural average temperature
for the planetary surface:

\begin{equation}
T_{ave}(t) = {1 \over 4 \pi r^2} \int T({\bf x}, t) d{\bf x}
\end{equation}

As Gerlich and Tscheuschner note\cite{gerlich} in their section 3.7, thanks to
H\"{o}lder's inequality, this average temperature $T_{ave}(t)$ is always
less than or equal to the effective thermal radiation temperature
$T_{eff}(t)$, so $T_{ave}^4$ is less than or equal to $T_{eff}^4$, and
rearranging Eq. \ref{planetNetEnergy} gives the following constraint on average
temperature.

\begin{equation}
T_{ave}(t)^4 \le {1 \over \sigma \epsilon_{eff}(t)} ((1 - a_{eff}(t)) S(t)/4
 - \dot{E}_{planet}(t)/4 \pi r^2)
\end{equation}

\section{Some Examples}
\subsection{Model 1: Nonrotating planet}
First let's look at the simple model planet solved by Gerlich and Tscheuschner
(section 3.7.4) This is a non-rotating planet (or a planet with a rotation
axis parallel to the incoming radiation) with no internal heat
transport in constant local radiative equilibrium so that $\dot{E}_{planet}$
is always zero. The non-rotation removes all time-dependences.
Emissivity is assumed to be 1 everywhere; likewise $S$
and $a$ are uniform. Also the microwave background is ignored
so the un-lit side of the planet is always at absolute zero
temperature. From Eq. \ref{incomingLocalIrradiance} for the
local irradiance $s({\bf x})$ we
quickly obtain the local temperature for the lit side of the planet:

\begin{equation}
T_{model 1}({\bf x}) = \{(1 - a) \cos(\theta({\bf x})) S/\sigma \}^{1/4}
\end{equation}

The average temperature is obtained by integrating over the sphere:
\begin{equation}
T_{ave} = {1 \over 4 \pi} ((1-a) S/\sigma)^{1/4}
    \int_0^{\pi/2} \cos(\theta)^{1/4} 2 \pi \sin(\theta) d\theta
   = {2 \over 5} ((1-a) S/\sigma)^{1/4}
\end{equation}

The effective temperature similarly is given by
\begin{equation}
T_{eff}^4 = {1 \over 4 \pi} ((1-a) S/\sigma)
    \int_0^{\pi/2} \cos(\theta) 2 \pi \sin(\theta) d\theta
   = {1 \over 4} ((1-a) S/\sigma)
\label{tempEffNonrotating}
\end{equation}

which is what it has to be to ensure (Eq. \ref{planetNetEnergy})
that $\dot{E}_{planet}$ is zero.

So in this case the ratio $T_{ave}/T_{eff}$ is $2\sqrt{2}/5$ or about 0.566,
and the planet's average temperature is indeed well below the
effective temperature in this simple model. Plugging in numbers appropriate
for Earth, $T_{eff}$ comes to 255 K and $T_{ave}$ would be 144 K, for this
non-rotating atmosphere-free version of the planet.

\subsection{Model 2: Simple rotating planet}

To this simple model let us now add rotation, including a local heat
capacity effect that accounts for some heat transport vertically, while still
leaving out any transport of heat horizontally from
one location to another. Assume the radiation direction is
in the plane of rotation. Define the rotation period $D$ (a day for the
planet) and a thermal inertia coefficient\cite{rtp} $c$ with units
of J/Km$^2$. On a real planet, $c$ depends on temperature
and on $D$ (a time- or frequency- dependence); physically it
represents the product of
the volumetric heat capacity and the depth or height to which the incident
heat energy is circulated or conducted during a daily thermal cycle. $c$ then
determines the local rate of change of temperature based on the local
version of the net energy equation:

\begin{equation}
c \dot{T}({\bf x}, t) = E_{absorbed}({\bf x}, t) - E_{emitted}({\bf x}, t)
	= (1 - a) S \cos(\theta({\bf x}, t))\Theta(\cos(\theta({\bf x}, t)))
       - \sigma T({\bf x}, t)^4   \label{TRotatingEqn}
\end{equation}

Represent ${\bf x}$ by angular coordinates $\phi$ for longitude and $\xi$
for latitude. Thanks to the rotation, the position of ${\bf x}$
relative to the incoming sunlight changes as if $\phi$ were steadily
incremented at a rate $2 \pi t/D$. The sun angle $\theta$ then is
found from:

\begin{equation}
\cos(\theta) = \cos(\phi + 2\pi t/D) \cos \xi
\end{equation}

The driving forces in the equation repeat with period $D$, so
under steady state conditions the solution(s) of Eq. \ref{TRotatingEqn} should
also repeat with that period. At any point $\phi, \xi$ on the surface
this solution would follow exactly the same curve of temperature as a function
of time for every longitude $\phi$, at the given latitude $\xi$. In the
following we replace $2\pi/D$ with the symbol $\omega$ and
$(1-a) S \cos \xi$ with $A$, and for simplicity set $\phi = -\pi/2$ (for all
other points on the surface the solution is just shifted forward or back
a bit in time). Then the step function $\Theta(\cos(\phi + \omega t))$ becomes
equivalent to a square wave $W(\omega t)$ which is $1$ for $\omega t$ between
$0$ and $\pi$, $0$ for $\omega t$ between $\pi$ and $2 \pi$, and repeating
periodically after that.

So Eq. \ref{TRotatingEqn} is reduced to:

\begin{equation}
c \dot{T} = A \sin(\omega t) W(\omega t) - \sigma T^4   \label{TRotatingSimp}
\end{equation}

Any non-transient solution for $T(t)$ will be periodic in time so that
$T(D) = T(0)$. Integrating Eq. \ref{TRotatingSimp} over a planetary day (t = 0
to t = D) gives:

\begin{equation}
0 = { 2 A \over \omega } - \sigma \int_0^D T^4 dt \label{integratedRot}
\end{equation}

Define an effective radiative temperature $T_{eff}(\xi)$ for latitude $\xi$
based on the average fourth power (whether averaged over time or over
longitudes is the same):

\begin{equation}
T_{eff}(\xi)^4 = {1 \over D} \int_0^D T^4 dt
\end{equation}

then rearranging Eq. \ref{integratedRot} and substituting in the definitions
of $\omega$ and $A$ gives:
\begin{equation}
T_{eff}(\xi)^4 = { (1-a) S \cos \xi \over  \pi \sigma } \label{localEffectiveTemp}
\end{equation}

Note that the peak $T_{eff}$ for $\xi = 0$ (the equator) is a factor of
$1/\pi^{1/4} \approx 0.75$ times the peak temperature on the non-rotating
planet (from the point directly under the sun).

Once again we can check that the rate of change in net energy for
the planet as a whole (Eq. \ref{planetNetEnergy}) comes to zero by
finding the effective radiative temperature for the entire surface:

\begin{equation}
T_{eff}^4 = {1 \over 4 \pi} \int_{-\pi/2}^{\pi/2}
 {(1 - a) S \cos \xi \over \pi \sigma} \cdot 2 \pi \cos \xi d\xi
 = {(1 - a) S \over 2 \pi \sigma} \int_{-\pi/2}^{\pi/2} \cos^2 \xi d\xi
 = {(1 -a) S \over 4 \sigma} \label{tempEffRotating}
\end{equation}

as it has to be, the same as for the non-rotating case in
Eq. \ref{tempEffNonrotating}.

While we won't find a full analytic form for the temperature as a function of
latitude and time in this model, we can learn a bit by examining the
time dependence of temperature in Eq. \ref{TRotatingSimp} more closely.
First, define $x = \omega t$ and $y(x) = T(\xi, t)/T_{eff}(\xi)$ for a
given latitude $\xi$, with $T_{eff}(\xi)$ determined by
Eq. \ref{localEffectiveTemp}. Eq. \ref{TRotatingSimp} then reduces to:

\begin{equation}
{dy \over dx} = {(1 - a) S \cos \xi \over c \; \omega T_{eff}(\xi)} \sin(x) W(x) -
  { \sigma T_{eff}(\xi)^3 \over c \; \omega } y^4
 = \lambda (\sin(x) W(x) - {1 \over \pi} y^4)
\label{odeForYSimple}
\end{equation}

where we define the dimensionless parameter
$\lambda = (1-a) S \cos \xi / c \; \omega T_{eff}(\xi)$
Physically this roughly represents the ratio of the quantity of incoming
energy absorbed in a day to the total heat content of the surface
(to a relevant depth) at the effective radiative temperature. If
$\lambda$ is small (heat capacity or rotation frequency high or latitude
close to the poles), heating or cooling will occur only slowly,
and the temperature will stay close to $T_{eff}$ throughout the day
($y$ will be close to $1$). If $\lambda$ is large (heat capacity or
rotation frequency low, latitude closer to the equator) then heating
and cooling are rapid, and the temperature variation is more significant.

\begin{figure}
\includegraphics[width=16cm]{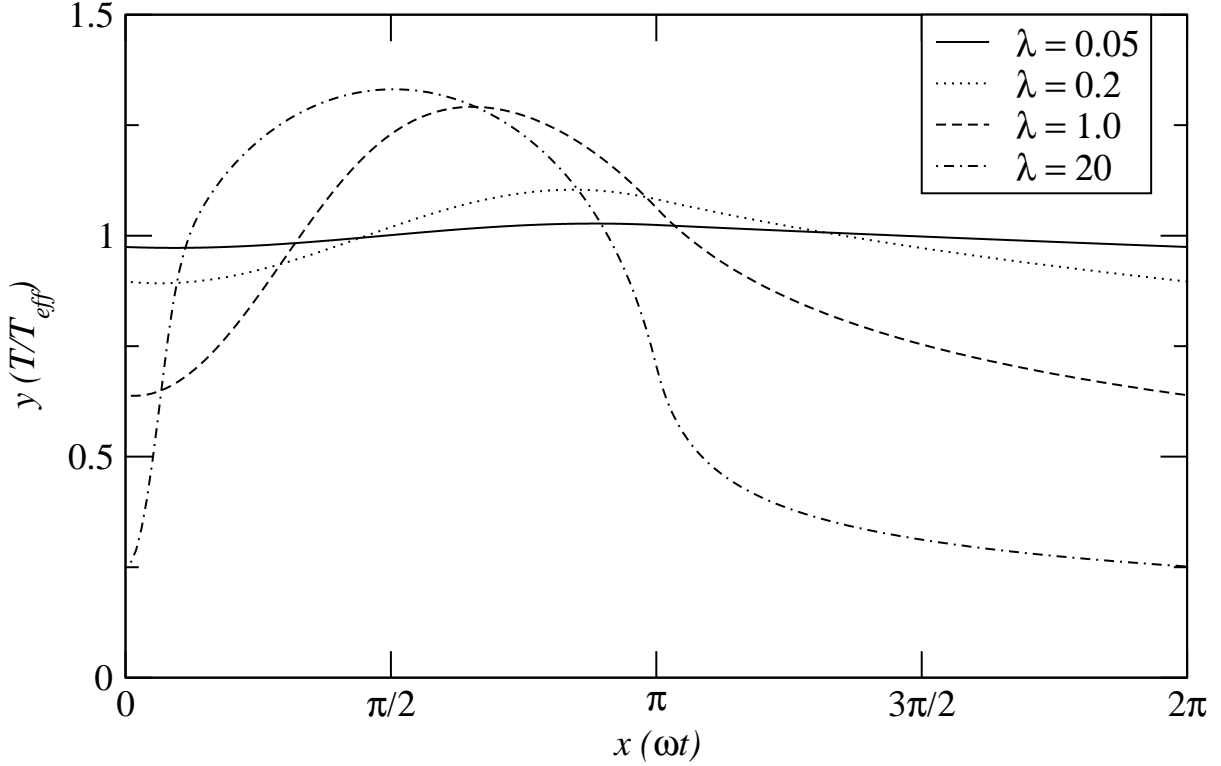}
\newline
\caption{Temperature relative to the effective radiating temperature
($T_{eff}$) for the simple rotating planet model for various
values of the thermal response parameter $\lambda$. The plots are of
temperature against time where $0$ is sunrise, $\pi$ is sunset,
and $2\pi$ is sunrise again.
\label{rotTemp}
}
\end{figure}

We can find an analytic solution for the night side of the planet, where
the square wave $W(x) = 0$ ($x$ between $(2n-1) \pi$ and $2n \pi$ for
integer $n$). Eq. \ref{odeForYSimple} loses all dependence on $x$ and is
easily integrated:

\begin{equation}
{dy \over dx} = -{\lambda \over \pi} y^4 \Rightarrow
y(x) = ({\pi \over 3 \lambda (x - a)})^{1/3}
\end{equation}

where $a$ is a constant determined by the initial condition $y(\pi)$
(the value of the temperature when night begins for $x = \pi$):

\begin{equation}
a = \pi (1 - {1 \over 3 \lambda y(\pi)^3}) \Rightarrow
y(2\pi)/y(\pi) = 1 / (1 + 3 \lambda y(\pi)^3)^{1/3}
\end{equation}

which gives us the night-time temperature drop. When $\lambda$ is small
and $y$ is not too large to start with, the change is small - a fractional
decline of roughly $\lambda y(\pi)^3$. For large values of $\lambda$,
the night-time temperature drop is limited by this slow
inverse $1/3$ power. This makes sense as the rate of temperature decrease
must decline sharply as temperature gets lower and the $T^4$ radiative
term drops.

For the daytime, if the temperature starts out low with $y << 1$, then
the $y^4$ term is negligible, at least at first, and Eq. \ref{odeForYSimple}
can be integrated easily enough:

\begin{equation}
{dy \over dx} \approx \lambda \sin x \Rightarrow
y(x) \approx b - \lambda \cos x
\end{equation}

where the inegration constant $b$ again is set by the initial
condition $b = y(0) + \lambda$, which $y(\pi) = y(0) + 2 \lambda$, i.e.
the temperature increments by $2 \lambda$ on a sinusoidal curve during
the day. Of course when $\lambda$ is large or $y(0)$ starts close to one,
this approximation breaks down.

Another approximation is to assume variations in $y$ are small and
that we can linearize about some chosen value $y_0$. This gives:

\begin{equation}
{dy \over dx} \approx \lambda \sin x - {\lambda y_0^4\over \pi} -
   4 {\lambda y_0^3 \over \pi} (y-y_0)
\end{equation}

which as a linear ordinary differential equation yields a solution:

\begin{equation}
y(x) = {3 \over 4} y_0 + \beta e^{-4 \lambda y_0^3 x/\pi} +
   {\lambda \over 1 + (4\lambda y_0^3/\pi)^2} ({4 \lambda y_0^3 \over \pi} \sin x - \cos x)
\end{equation}

$\beta$ here is another constant of integration to be determined by
an appropriate initial condition. The linearization fails once
$y$ deviates significantly from $y_0$, but the result should be generally
valid if $\lambda$ is small, and can be used to generate
step-wise solutions for the daytime temperatures under any value of $\lambda$;
numerical intergration of the basic equation of course can do the same.

\begin{figure}
\includegraphics[width=16cm]{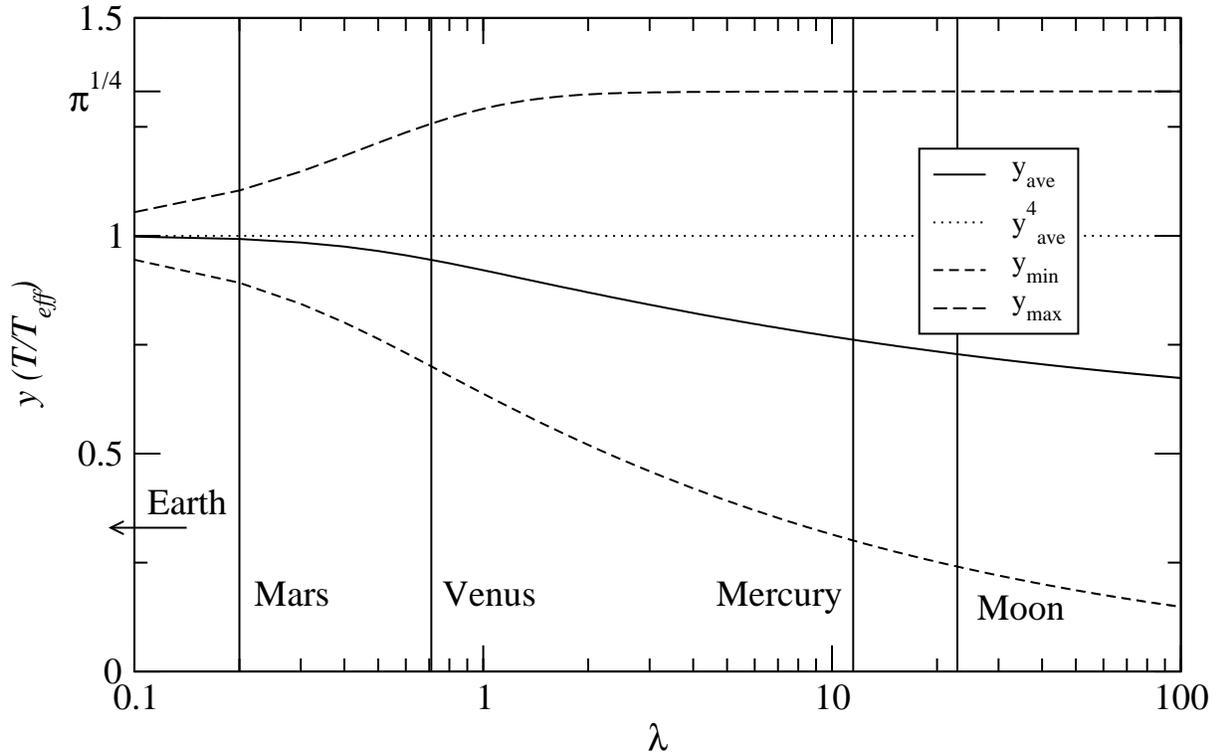}
\newline
\caption{The average, minimum, and maximum values of the relative
temperature $y$ from numerically integrating the equations for
different values of $\lambda$. The numerically computed average
value of $y^4$ is also shown; this should always be exactly 1.
Lines are shown indicating the approximate values of
$\lambda$ corresponding to the equators of the terrestrial planets,
which gives a picture of their most extreme temperature profiles
if they had infrared-transparent atmospheres.
\label{rotAll}
}
\end{figure}

Numerical computed solutions for various values of $\lambda$ are shown
in Fig. \ref{rotTemp}; Fig. \ref{rotAll} shows the
trends for average $y$ ($T/T_{eff}$),
average fourth power of $y$, and the minimum and maximum $y$ values as
$\lambda$ increases. As expected, the average fourth power is fixed
at 1, while the average $y$ decreases as $\lambda$ increases, eventually
approaching the non-rotating value of $2\sqrt{2}/5$ as
$\lambda \rightarrow \infty$. Also note the maximum temperature quickly
approaches the nonrotating value of $y = \sqrt[4]{\pi}$ for large $\lambda$.
The minimum temperature drops slowly as $y \sim 1/(3\lambda)^{1/3}$.

\begin{table}[b]
\caption{Relevant parameters for the planets. See
http://nssdc.gsfc.nasa.gov/planetary/factsheet/. $\lambda$ for
Eq. \ref{odeForYSimple} (at the equator, $\xi = 0$) estimated from thermal
inertia, solar day, and the other parameters. It is particularly small for
Earth thanks to rapid rotation and the high heat capacity of water
covering most of the surface.}
\begin{tabular}{c c c c c c c c}
\hline
Planet & solar constant & albedo & solar day & $T_{eff}$ & $T_{ave}$ & Difference & $\lambda$ \\
 & (W/m$^2$) & & (Earth days) & (K) & (K) & (K) & \\
\hline
Mercury & 9127 & 0.12 & 176 & 434 & ? & ? &  11 \\
Venus & 2615 & 0.75 & 117 & 232 & 737 & 505 & 0.7 \\
Earth & 1367 & 0.306 & 1 & 255 & 288 & 33 &  0.04 \\
Moon & 1367 & 0.11 & 29.53 & 270 & 253 & -17 & 20 \\
Mars & 589 & 0.25 & 1.03 & 210 & 210 & 0 & 0.2 \\
\end{tabular}
\end{table}

Approximate formulas for the average temperature are, for large $\lambda$:
\begin{equation}
y_{ave} \sim 2\sqrt{2}/5 + 0.392 \lambda^{-0.279} ; \lambda \rightarrow \infty
\end{equation}
and for small $\lambda$:
\begin{equation}
y_{ave} \sim 1 - 0.196 \lambda^2 ; \lambda \rightarrow 0
\label{aveInSmallLambdaLimit}
\end{equation}

Note that from Eq. \ref{localEffectiveTemp} the temperature scale varies
with latitude as $\cos(\xi)^{1/4}$, while the value of $\lambda$
varies as $\cos(\xi)^{3/4}$. So finally, integrating over the whole
planet we have an average temperature value of:

\begin{equation}
T_{ave} = ({(1-a) S \over \pi\sigma})^{1/4}
    \int_0^{\pi/2} \cos(\xi)^{1/4} y_{ave}(\lambda(\xi)) \cos(\xi) d\xi
\label{aveTempForRotating}
\end{equation}

For small $\lambda$ we could substitute the expression from
Eq. \ref{aveInSmallLambdaLimit}; in any case, we know $y_{ave} < 1$ for
all latitudes, so we have an upper bound on the average temperature
of the entire rotating planet $T_{ave}$ by substituting
in the numerical value for the $\cos(\xi)^{5/4}$ integral:

\begin{equation}
T_{ave} < 0.69921 ({(1-a) S \over \sigma})^{1/4}
\end{equation}

and note that this bound is a little over 1\% less
than $T_{eff}$ for the entire planet
(Eq. \ref{tempEffRotating}) which has a
constant $(1/4)^{1/4} = 0.7071...$ instead of 0.69921 in the same expression.

So no matter the rotation rate, no matter the surface heat capacity,
the average temperature of the planet in this rotating example, with
only radiative energy flows and no absorbing layer in the atmosphere,
is always less than the effective radiating temperature. For
very slow rotation or low heat capacity it can be significantly less;
for parameters in the other direction it can come as close as 1\%
(i.e. up to 252 K on a planet like Earth).

\subsection{Model 3: Rotating planet with varying albedo}
While the variability in infrared emissivity is relatively small
across the surface of a realistic planet, the albedo can be
significantly different from place to place. One of these involves
taking into account the effect of ice, by which the high latitudes
reflect more incoming radiation back into space than equatorial
latitudes do. What effect does this have on effective radiating
temperature and total temperature?

We can model this by a slight change in model 2, by making the value
of $a$ dependent on latitude $\xi$. For example
let $a = \sin^2(\xi)$ so it is zero
at the equator, and approaches 1 at the poles.
This changes nothing in most of the
analysis of the preceding section, until we integrate over latitudes.
For Eq. \ref{tempEffRotating} we now have:

\begin{equation}
T_{eff}^4 = {1 \over 4 \pi} \int_{-\pi/2}^{\pi/2}
 {(1 - a) S \cos \xi \over \pi \sigma} \cdot 2 \pi \cos \xi d\xi
 = {S \over \pi \sigma} \int_{0}^{\pi/2} \cos^4 \xi d\xi
 = {3 S \over 16 \sigma} \label{tempEffVarAlbedo}
\end{equation}

This just means that our $\sin^2$ albedo has the same effect on the
total radiation absorbed by the planet as would a uniform albedo
value of $1/4$. However, it redistributes that energy, putting more
near the equator and less near the poles. The effect on average
temperature across the planet, for this modified version
of Eq. \ref{aveTempForRotating} is:

\begin{equation}
T_{ave} = ({S \over \pi\sigma})^{1/4}
    \int_0^{\pi/2} \cos(\xi)^{3/4} y_{ave}(\lambda(\xi)) \cos(\xi) d\xi
\label{aveTempForVarAlbedo}
\end{equation}

which then, putting in the numerical value for the $\cos^{7/4}$ integral,
gives the inequality

\begin{equation}
T_{ave} < 0.6206 ({S \over \sigma})^{1/4}
\end{equation}

which is about 5\% below the effective temperature (the numerical
coefficient is $(3/16)^{1/4}$ or $0.6580$).

\section{Infrared Absorption in the Atmosphere}
The examples of these simple models show that vertical energy transport
for a planet with a transparent atmosphere only smooths out the
daily temperature curve, without being able to bring the surface
temperature higher than the effective radiative temperature. The
same is true if we were to add in more realistic horizontal
energy transport from larger-scale atmospheric and oceanic
circulation - of course getting much more realistic means entering
the realm of more full-scale general circulation models\cite{weart}, which
we have no intention of doing here.

On a planet with significant internal energy sources the effective
temperature for radiative balance could be exceeded even with
a transparent atmosphere. For example a planet still losing its
initial heat of formation, or a planet remote from its sun
with a high enough radioactive content, or on a planet or moon with
very large tidal forces, you will have a net outward flow of energy
to space, and may well have an average temperature above the limit.
But for the terrestrial worlds of our solar system, these internal
sources of heat are thousands of times too small to have any noticeable
effect on surface temperature.

And yet the observed average surface temperature on Earth and Venus
significantly exceeds the effective radiative temperature set
by the incoming solar radiation. This is not observed for the Moon or
Mars. What makes Venus and Earth so different?

Net energy flux is determined by the radiation that gets into space,
not what leaves the surface. The only way for a planet to
be radiatively warmer than the incoming sunlight allows is for some
of that thermal radiation to be blocked from leaving. That means
some layer above the surface must be absorbing or reflecting
a significant fraction of the outgoing infrared radiation. I.e.
the atmosphere must not be transparent to infrared.

So, let's add to our rotating planet model a simple model
of this blocking effect: a fraction $f$ (between 0 and 1) of the
outgoing radiation $E_{emitted}$ from the surface is absorbed by a thin layer
of the atmosphere. This layer will have its own temperature but for
simplicity we make the assumption that the heat capacity of the
atmospheric layer is low so that it remains essentially
radiatively balanced through the day, and the specific temperature becomes
irrelevant. That means that this atmospheric layer continuously emits
an amount $f \cdot E_{emitted}$ equal to what it absorbs from the ground.

Since thermal re-emission is randomly directed, half the radiation
from this atmospheric layer will go up, and half down. Assuming the surface
is fully absorbing and the rest of the atmosphere is transparent, total
outgoing radiation from the planet (above the atmospheric layer) is then:

\begin{equation}
E_{out} = (1 - f) E_{emitted} + {1 \over 2} f E_{emitted}
      = (1 - f/2) E_{emitted}
\end{equation}

while absorbed radiation on the surface is now:
\begin{equation}
E_{absorbed}(t) = \pi r^2 (1 - a_{eff}(t)) S(t) + {1 \over 2} f E_{emitted}
\end{equation}

Incoming solar radiation still drives everything - if the solar
constant S drops, then so does everything else. But the effect of
the absorbing layer is to reduce the final outgoing energy for
a given temperature, so the planet heats up until things are
back in balance again.

Generalizing Eq. \ref{planetNetEnergy} we have net energy change
(which can be calculated either at the surface or above the absorbing
layer of the atmosphere):

\begin{eqnarray}
\dot{E}_{surface}(t) = E_{absorbed}(t) - E_{emitted}(t) =
 \pi r^2 (1 - a_{eff}(t)) S(t) + {1 \over 2} f E_{emitted}(t)
   - 4 \pi r^2 \sigma \epsilon_{eff}(t) T_{eff}(t)^4 \\
 = \pi r^2 (1 - a_{eff}(t)) S(t) -
    4\pi r^2 \sigma \epsilon_{eff}(t) (1 - f/2) T_{eff}(t)^4
\label{planetNetEnergyWithAbs}
\end{eqnarray}

We then end up with essentially the same equations as in
the previous section, for example Eq. \ref{TRotatingEqn} is the
same, except that effectively the solar input $S$ and thermal inertia
$c$ in those equations are increased by the factor $1/(1-f/2)$.

That means the surface effective radiative temperature $T_{eff}$ in
those equations is increased by a factor $(1/(1-f/2))^{1/4}$, or as
much as $2^{1/4}$ for a fully absorbing atmospheric layer. The
parameter $\lambda$ is then reduced by that same ratio (the increases in
$S$ and $c$ cancel out, leaving a $1/T_{eff}$ term). So the
temperature curve of this radiatively insulated planet is even
more uniform than without the insulating layer. The average temperature
can come within a few percent of this higher $T_{eff}$, or well
above the limits for a planet with a transparent atmosphere.

A more realistic atmosphere would be characterized by more than one
absorbing layer (or a thick layer with a temperature differential
and limited conductivity from bottom to top), which will further
decrease outgoing thermal radiation and increase surface temperatures.
Details of absorption in the real atmosphere also depend on pressure;
nevertheless, the presence of any absorption at all is what qualitatively
distinguishes a greenhouse-effect planet from one with a transparent
atmosphere, and is what allows surface temperatures to climb
above the effective radiative limit.

\section{Conclusion}
Gerlich and Tscheuschner\cite{gerlich} state, among more
extravagant claims, that ``Unfortunately, there is no source in the
literature, where the greenhouse effect is introduced in harmony with
the scientific standards of theoretical physics.'' The above
analysis I believe completely establishes, within perfectly
simple and appropriate theoretical physics constructs, the main
points. Namely that assuming ``the atmosphere is transparent for
visible light but opaque for infrared radiation'' leads to ``a warming
of the Earth's surface'' relative to firm limits established
by basic physical principles of energy conservation, for the
case of an atmosphere transparent to both visible and infrared.

In particular, it has been shown that:

\begin{enumerate}
\item An average surface temperature
for a planet is perfectly well defined with or without rotation,
and with or without infrared absorbing gases

\item This average temperature is mathematically constrained to be less
than the fourth root of the average fourth power of the temperature, and
can in some circumstances (a planet with no or very slow rotation,
and low surface thermal inertia) be much less

\item For a planet with no infrared absorbing or reflecting layer
above the surface (and no significant flux of internal energy),
the fourth power of the surface temperature
always eventually averages to a value determined by the incoming
stellar energy flux and relevant reflectivity and emissivity parameters.

\item The only way the fourth power of the surface temperature can
exceed this limit is to be covered by an atmosphere that is at least
partially opaque to infrared radiation. This is the atmospheric
greenhouse effect.

\item The measured average temperature of Earth's surface
is 33 degrees C higher than the limit determined by items (2) and (3).
Therefore, Earth is proved to have a greenhouse effect of at least
33K.

\end{enumerate}

The specific contributions of individual gases such as CO$_2$ to
Earth's greenhouse effect are covered well by the standard
treatments of the subject\cite{liou,rtp,weart}.

\end{document}